\documentclass{article}
\usepackage[utf8]{inputenc}
\usepackage{authblk}
\usepackage{multirow}
\usepackage{amsfonts,amsmath,amssymb,amsthm,mathtools}
\usepackage{algorithm}
\usepackage{algpseudocode}
\usepackage{graphicx} % 用于 \scalebox
\usepackage[table]{xcolor} % 启用表格颜色支持
\usepackage{subcaption} % 用于创建子图和对应的标题
\usepackage{listings}
\usepackage{listings}
\usepackage{xcolor}
\usepackage{setspace}
\usepackage[margin=1.25in]{geometry}
\usepackage{graphicx}
\graphicspath{ {./figures/} }
\usepackage{subcaption}
\usepackage{booktabs}
\usepackage{lineno}
\usepackage{hyperref}
\title{From Understanding to Excelling: Template-Free Algorithm Design through Structural-Functional Co-Evolution} 

\author[1,2]{Zhe Zhao}
\author[5]{Haibin Wen}
\author[1]{Pengkun Wang}
\author[2]{Ye Wei}
\author[3]{Zaixi Zhang}
\author[2]{Xi Lin}
\author[2]{Fei Liu}
\author[4]{Bo An}
\author[5]{Hui Xiong}
\author[1]{Yang Wang}
\author[2]{Qingfu Zhang}

\affil[1]{University of Science and Technology of China, Hefei 230026, China}
\affil[2]{City University of Hong Kong, Hong Kong 999077, China}
\affil[3]{Princeton University, New Jersey 08544, USA}
\affil[4]{Nanyang Technological University, Singapore 639798, Singapore}
\affil[5]{The Hong Kong University of Science and Technology (Guangzhou), Guangzhou 511458, China}

\date{}

\begin{document}

\maketitle

%%%%%% Abstract %%%%%%
\begin{abstract}
Large Language Models (LLMs) have significantly accelerated the automation of algorithm generation and optimization. However, existing methods (e.g., EoH and FunSearch) mainly rely on predefined templates and manual calibration, focusing only on the local evolution of key functions pre-identified by humans. As a result, these approaches struggle to fully exploit the synergistic effects of the overall architecture and realize the full potential of global optimization.

In this paper, we propose an end-to-end algorithm generation and optimization framework based on LLMs. The core idea is to leverage the deep semantic understanding and information extraction capabilities of LLMs to transform natural language requirements into code solutions, combined with a bi-dimensional co-evolution strategy that concurrently optimizes both the functional and structural aspects of the algorithm. This closed-loop, automated design process spans from problem analysis and code generation to global optimization. Our framework not only overcomes the limitations imposed by fixed templates but also automatically identifies key algorithmic modules to perform multi-level joint optimization, thereby continuously enhancing the performance and innovative structural design of the generated code.

Extensive experiments conducted on various algorithm design scenarios, including the Traveling Salesman Problem and large-scale optimization tasks, demonstrate that our approach significantly outperforms traditional local optimization methods in terms of both performance and innovation. Further studies reveal its high adaptability to unknown environments and its breakthrough potential in structural design. Experimental results indicate that our method can generate and optimize innovative algorithms that surpass human expert designs, thus broadening the applicability of LLMs for algorithm design and offering a novel solution pathway for automated algorithm development.
\end{abstract}

\section{Introduction}

Algorithm design is a core challenge in computer science, playing a critical role in driving scientific discovery, engineering innovation, and solving computational problems \cite{fei2024eoh, mendling2023methodology, FunSearch2023, ye2024reevo, kerschke2019automated, lake2017building}. Algorithms serve as the underlying engine of modern systems, powering optimization, decision-making, and learning across diverse fields such as bioinformatics, robotics, and finance \cite{cormen2009introduction, pearl1984heuristics, skiena2020algorithm, sedgewick2011algorithms, knuth1997art}. Nonetheless, designing efficient algorithms for complex problems remains a labor-intensive process that demands extensive domain expertise, iterative manual refinement, and a deep understanding of algorithmic paradigms \cite{hoos2004stochastic, garey1979computers, karp1972reducibility, wolpert1997no, mitchell1998introduction}. These challenges severely limit the scalability of algorithm development and impede the exploration of unconventional or truly innovative designs. Achieving end-to-end automation—from problem specification to generating optimized solutions—not only promises to accelerate innovation and reduce human intervention but also holds the potential to uncover designs beyond current human intuition \cite{zhao2023autooptlib, yu2024deep, real2020automl, elsken2019neural, feurer2015efficient}.

Recent advances in evolutionary computation and metaheuristic optimization have partially automated parameter tuning and heuristic development. More recently, machine learning methods have enabled systems to learn optimization strategies from data and synthesize code snippets tailored to specific tasks \cite{vanstein2024intheloop, Stanovov2022NeuroevolutionFP, Vladimir2024AutomaticDO}. In particular, large language models (LLMs) such as GPT have emerged as transformative tools, capable of mapping natural language requirements directly into executable code. Building on these breakthroughs, approaches such as \textit{EOH} \cite{fei2024eoh} and \textit{Reevo} \cite{ye2024reevo} have integrated evolutionary computation and machine learning to automate algorithm design. Despite these advancements, current methods primarily operate within fixed algorithmic templates and focus on local iterative optimizations. This leads to shortsighted improvements, as they overlook the collaborative potential of global architectural innovation and heavily rely on manually constrained search spaces \cite{liu2024, musslick2024automating}.

These inherent limitations expose a significant gap in achieving truly end-to-end automation in algorithm design. While current systems can refine known strategies, they typically fall short in exploring new paradigms or reimagining algorithm structures. To overcome these constraints, any automated framework must transcend traditional function-level optimizations by adopting a holistic design perspective—integrating structural innovation with global optimization.

To address these challenges, this paper introduces a novel framework that integrates the capabilities of evolutionary computation, machine learning, and large language models into a unified end-to-end system. Our method emphasizes global design exploration and structural innovation, moving beyond local function-level fine-tuning to optimize entire algorithm architectures. The main contributions of this work can be summarized as follows:

\begin{itemize}
    \item \textbf{Deep Understanding of Algorithm Requirements and Template-Free Generation:} Our system leverages information from input papers and knowledge bases to automatically construct a complete workflow, thereby achieving a holistic understanding of algorithmic requirements. On this foundation, the system autonomously explores innovative architectures and uncovers non-intuitive design paradigms while completely discarding predefined algorithmic templates. This innovative paradigm breaks traditional design constraints, enabling the generation of breakthrough and highly innovative algorithms.
    
    \item \textbf{Holistic Architecture Optimization and Autonomous Search Space Expansion:} By globally optimizing the entire algorithm architecture, our approach precisely captures and harnesses the intricate interactions among components, achieving significant breakthroughs in overall performance and design. Simultaneously, the framework dynamically defines and expands its own search space, greatly reducing reliance on manually specified constraints and prior domain knowledge, thereby enhancing the system's adaptability and innovation in novel scenarios.
\end{itemize}

\begin{figure*}[ht]
\centering
\includegraphics[width=1\textwidth]{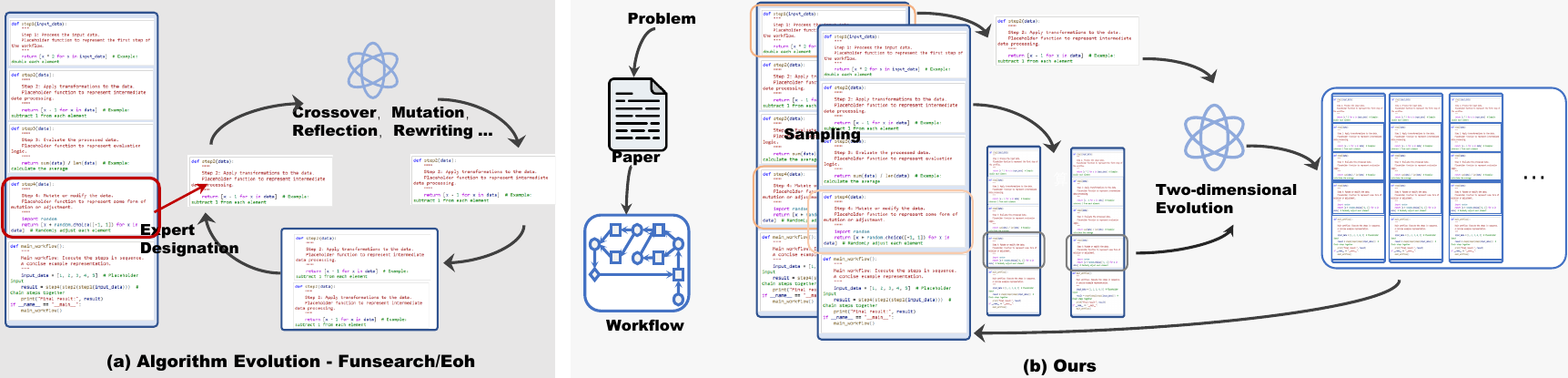}
\caption{\textbf{Schematic Comparison of Algorithm Evolution}:\textbf{ (a)} Methods such as FunSearch/EOH primarily rely on predefined templates and iterate improvements at the level of individual functional modules through operations like crossover, mutation, reflection, and rewriting;\textbf{ (b) }The proposed method leverages the deep semantic parsing capabilities of large language models to build a complete workflow, enabling multi-dimensional collaborative evolution of both algorithm structure and function. This approach breaks the limitations of single-function or local operations, achieving autonomous exploration and global optimization of the entire architecture, and providing adaptive, innovative solutions for complex problems.}
\label{fig1}
\end{figure*}

The proposed framework not only alleviates the limitations of existing technologies but also paves the way for fully automated and innovative algorithm design. By significantly lowering the threshold for high-performance algorithm development, this work has the potential to expedite solutions to complex computational challenges and ignite new research directions in computational science. The remainder of this paper is organized as follows: Section \ref{sec2} reviews related work; Section \ref{sec3} details the proposed method; Section \ref{sec4} presents extensive experimental validations; Section \ref{sec5} provides a comprehensive case study demonstrating end-to-end automation and superior performance relative to human-designed methods; 
% Section \ref{sec6} presents other cases that further demonstrate the excellent performance of our LLM-driven automated algorithm design approach in solving complex problems; 
and finally, Sections \ref{sec7} and \ref{sec8} conclude the paper with discussions on future research directions.

\section{Related Work}\label{sec2}
\subsection{Traditional Hyper-Heuristic Approaches (HH)}
Traditional hyper-heuristic (HH) methods have long been considered a promising approach to achieving automation in algorithm design by abstracting the process of heuristic selection and combination \cite{burke2003hyper}. These methods typically operate by either selecting the most effective heuristics from a predefined set or combining simpler components to create new heuristics \cite{chakhlevitch2008hyperheuristics}. This abstraction enables HH to be applied to a wide range of optimization problems, offering greater generality compared to problem-specific methods \cite{ozcan2008comprehensive}. However, the reliance on predefined heuristic spaces—often manually crafted by domain experts \cite{kerschke2019automated}—represents a fundamental limitation, restricting their ability to explore innovative algorithmic structures or paradigms.

Traditional HH methods primarily focus on parameter tuning and improving existing heuristics, rather than designing entirely new algorithms \cite{blum2003metaheuristics}. This focus stems from their dependence on manually defined search spaces, which limits their ability to discover new approaches that may outperform traditional heuristics \cite{o2009riccardo}. For example, HH frameworks based on genetic programming \cite{stanovov2022neuroevolution} often require significant manual intervention to define reusable algorithm components, thereby constraining their scalability and adaptability to complex, dynamic problems. Moreover, achieving true end-to-end automation in algorithm design remains a major challenge for traditional HH methods. These approaches often lack the capability for autonomous exploration and innovation beyond the constraints of manually defined templates. Consequently, their optimization processes are prone to early convergence to local optima and frequently fail to uncover unconventional paradigms that could lead to breakthroughs in algorithm design.
\subsection{LLMs for Algorithm Design}

The emergence of large language models (LLMs) has opened unprecedented opportunities for automating algorithm design. Their code generation capabilities have been successfully applied in various fields, including debugging, programming competitions, and optimization problem solving \cite{guo2023towards,liu2024evolution,brahmachary2024large}. Trained on extensive code datasets, LLMs not only generate syntactically correct code snippets but also efficiently tackle complex programming tasks using templates or zero-/few-shot prompts. However, current studies primarily focus on syntactic correctness and functional implementation—executing human instructions—thus falling significantly short of substituting expert-driven algorithm innovation.

EoH \cite{fei2024eoh} innovatively integrates LLMs with evolutionary algorithms to achieve, for the first time, function-level innovation in algorithmic components. In contrast to the later-published FunSearch \cite{romera2024mathematical}, EoH distinguishes itself by evolving not only the code of functional modules but also the underlying heuristic strategies. These approaches \cite{liu2024evolution,yao2024multi,yao2024evolve,wang2024llm,liu2024systematic} mark significant advancements in automating algorithm design by uncovering novel heuristics that outperform human-crafted methods in specific domains. Nonetheless, these methods remain heavily dependent on predefined templates, lack the capacity for autonomous exploration and innovation, and tend to focus on isolated components rather than holistic optimization—limitations that constrain the overall potential of automated algorithm design.

\section{Method}\label{sec3}
This section presents in detail an end-to-end code generation and optimization framework based on large language models. The framework formalizes the process of automatically transforming natural language problem descriptions into high-quality code solutions, and continuously improves code performance by adopting an adaptive generation mechanism combined with a bi-dimensional co-evolution strategy. The entire process is highly consistent with the modular workflow shown in Figure~\ref{fig1}: from problem analysis, code generation, and validation testing, to iterative optimization based on a feedback loop, progressively realizing breakthroughs at the global architectural level.

\subsection{Problem Definition and Formal Modeling}
In order to automatically transform natural language requirements into code solutions, the system first performs a rigorous formal modeling of the problem. Specifically, we abstract the code generation and optimization problem as a triplet optimization model:
\begin{equation}
    \mathcal{Q} = (\mathcal{R}, \mathcal{O}, \mathcal{C}),
\end{equation}
where $\mathcal{R}$ represents the functional requirements, $\mathcal{O}$ denotes the set of target performance metrics (such as execution efficiency and resource utilization), and $\mathcal{C}$ describes the constraints that must be satisfied in the design.

A specific code implementation can be represented as:
\begin{equation}
    \mathcal{W} = (F, R),
\end{equation}
where $F = \{f_1, f_2, \dots, f_m\}$ is the set of function modules constituting the code system, and $R$ characterizes the dependencies among these modules. To quantitatively evaluate the candidate solutions, we introduce a set of validation rules:
\begin{equation}
    \mathcal{V} = \{v_1, v_2, \dots, v_k\}.
\end{equation}

Based on the above model, the goal of our method is to select, from the search space $\mathcal{S}$ of all candidate solutions, the code solution that both satisfies all validation rules and is optimal with respect to the objective function:
\begin{equation}
    \mathcal{W}^* = \arg\min_{\mathcal{W}\in\mathcal{S}} \mathcal{O}(\mathcal{W}), \quad \text{subject to } \forall\, v_i \in \mathcal{V}, \; \mathcal{W} \text{ satisfies } v_i.
\end{equation}
This formal model provides a rigorous quantitative basis for subsequent code generation, validation, and evolutionary optimization, fundamentally transforming the entire automation process into a mathematical problem that can be solved using optimization algorithms.

\subsection{Adaptive Code Generation and Optimization Process}
Based on the above formal modeling, we have designed an adaptive generation and optimization process that fully leverages the deep semantic understanding and code generation capabilities of large language models (LLMs) and refines the generated results through real-time feedback. The entire process consists of three main stages, which correspond to the "problem analysis," "code generation," "validation," and "iterative optimization" modules shown in Figure~\ref{fig1}.

\subsubsection{Semantic Understanding Stage}
In the first stage, the system employs a language model $\mathcal{L}$ to expand and structure the input problem $\mathcal{Q}$ described in natural language:
\begin{equation}
    S \gets \mathcal{L}(\mathcal{Q}, P_s),
\end{equation}
where $P_s$ is a carefully designed prompt template. This stage aims to thoroughly extract and systematically organize the original functional requirements while identifying the optimization objectives $\mathcal{O}$ and constraints $\mathcal{C}$, thereby providing accurate semantic information for subsequent code generation and optimization.

\subsubsection{Code Generation Stage}
Based on the structured semantic representation $S$, the system combines a prior knowledge base $\mathcal{K}$ and a code generation prompt template $P_w$ to automatically generate an initial code solution:
\begin{equation}
    \mathcal{W},\, P_f \gets (F, R), \quad \text{where } P_f \gets \mathcal{L}(S, \mathcal{K}, P_w).
\end{equation}
The code solution $\mathcal{W}$ generated at this stage includes a set of function modules $F$ and the dependencies $R$ among them. The fusion strategy of multiple information sources ensures both the diversity and the high potential quality of the solution, providing an initial population for subsequent validation and evolutionary operations.

\subsubsection{Validation and Feedback Optimization Stage}
To ensure that the generated code solution meets practical functional and performance requirements, the system introduces a comprehensive validation mechanism. Specifically, the generated code solution is evaluated against the predefined set of validation rules $\mathcal{V}$:
\begin{equation}
    \mathcal{Z} \gets (\mathcal{W}, \mathcal{V}).
\end{equation}
The validation result $\mathcal{Z}$ quantifies the correctness and performance of the code solution. If the code solution fails to pass validation, the system utilizes the feedback information to automatically adjust and re-optimize the code, forming an adaptive process based on a validation-feedback loop until all candidate solutions satisfy the predefined rules or an iteration limit is reached. Finally, the system outputs an initial population composed of diverse code solutions:
\begin{equation}
    \mathcal{I} \gets \{F_1, F_2, \dots, F_n\},
\end{equation}
which lays the foundation for subsequent global optimization exploration.

\subsection{Bi-dimensional Co-Evolution Optimization Framework}
To achieve breakthroughs at the global architectural level, our method further introduces a bi-dimensional co-evolution strategy based on the initial population by jointly optimizing both the functional and structural dimensions. This is the key aspect that distinguishes our approach from traditional methods (e.g., FunSearch/EOH) and corresponds closely with the process shown in Figure~\ref{fig1}(b).

\subsubsection{Functional Dimension Optimization}
In the functional dimension, the function modules within each candidate are optimized independently. Specific operations such as short-term reflection, crossover, and mutation are employed to locally improve the function modules. Through the reflection module, which generates improvement suggestions for local implementations, the system is able to automatically optimize the performance of each individual function in iterative refinement, achieving gradual and meticulous enhancement.

\subsubsection{Structural Dimension Optimization}
In the structural dimension, the system is not limited to local function improvements but instead explores new algorithmic architectures through cross-individual function combination and rewriting. Specifically, by selecting function modules from different candidates in the population and constructing new module combinations based on expert system prompts and fusion suggestions generated by the large language model, the search space is continually expanded, thereby uncovering entirely new design paradigms that surpass traditional templates.

\subsubsection{Feedback Loop and Global Optimization}
Through the co-evolution of the functional and structural dimensions, our method establishes a global feedback loop system. The long-term reflection mechanism integrates information from each iteration—taking into account both the current state and historical optimization experience—so that the entire evolutionary process gradually converges toward the globally optimal code solution:
\begin{equation}
    \mathcal{I}^{(t+1)} = \text{Evaluate}\Bigl(\mathcal{I}^{(t)} \cup \mathcal{I}^{F}_{\text{new}} \cup \mathcal{I}^{S}_{\text{new}}, \mathcal{Z}\Bigr).
\end{equation}
Here, $\mathcal{I}^{F}_{\text{new}}$ denotes the new candidate solutions generated from the functional dimension, while $\mathcal{I}^{S}_{\text{new}}$ represents those generated from the structural dimension. Through continuous iterations, the system gradually approaches the code solution $\mathcal{W}^*$ that satisfies all validation rules and is optimal in overall performance.

In summary, our method achieves full-chain automation from unstructured natural language to optimal code implementation. Through template-free generation and a bi-dimensional co-evolution strategy, it not only ensures meticulous optimization of local modules but also promotes innovative breakthroughs in the overall architecture, providing a novel technical pathway for high-quality automated algorithm design for complex problems.

\section{Comparative Analysis with Baseline Methods}\label{sec4}

In this section, we present a comprehensive evaluation of the proposed CAE framework against state-of-the-art heuristic methods, including GHPP\cite{duflo19gp} and ReEvo, across multiple benchmark instances. The evaluation focuses on the optimality gap (\textit{Gap}), a key performance metric for quantifying the quality of solutions. The \textit{Gap} is computed as follows:

\begin{equation}
\text{Gap} = \frac{\text{Base Obj} - \text{CAE Obj}}{\text{Base Obj}} \times 100\%,
\end{equation}

where:
\begin{itemize}
    \item \textbf{Base Obj} refers to the objective value of the baseline method (e.g., GHPP ).
    \item \textbf{CAE Obj} corresponds to the objective value obtained by the proposed CAE method.
\end{itemize}

The experimental setup includes evaluations on well-known instances of the Traveling Salesman Problem (TSP) and the Capacitated Vehicle Routing Problem (CVRP). For each instance, the results are averaged over three independent runs, ensuring robustness and consistency. Below, we analyze the performance of CAE in comparison to the baseline methods.
\begin{table*}[t!]
    \caption{The performance of different heuristic algorithms was compared. We report the average optimality gap for each instance, where the baseline results are drawn from [15], and all results are averaged over 3 runs with different starting nodes each time.}
    \label{tab:comparisons}
    \scalebox{0.9}{
        \centering
        \begin{tabular}{l|ccc|l|ccc|l|ccc}
            \toprule
            Instance     & GHPP    & ReEvo     & CAE & Instance        & GHPP    & ReEvo    & CAE & Instance       & GHPP  & ReEvo    & CAE \\
            \midrule
            ts225        & 7.7     & \underline{6.6}      &  \textbf{4.6}       &eil51           & 10.2    & \underline{6.5}      &   \textbf{3.5}    &d657           & 16.3  & \underline{16.0}     &   \textbf{14.8}  \\
            rat99        & 14.1    & \underline{12.4}     &  \textbf{11.7}      &d493            & 15.6    & \underline{13.4}     &   \textbf{10.6}   &kroA150        & 15.6  & \underline{11.6}     &   \textbf{10.1}\\
            rl1889       & 21.1    & \underline{17.5}     &  \textbf{15.8}      &kroB100         & 14.1    & \underline{12.2}     &   \textbf{7.0}     &fl1577         & 17.6  & \underline{12.1}     &   \textbf{9.8}\\
            u1817        & 21.2    & \underline{16.6}     &  \textbf{13.0}      &kroC100         & 16.2    & \underline{15.9}     &   \textbf{6.8}     &u724           & 15.5  & \underline{16.9}     &   \textbf{15.1} \\
            d1655        & 18.7    & \underline{17.5}     &  \textbf{12.5}      &ch130           & 14.8    & \underline{9.4}      &   \textbf{7.8}     &pr264          & 24.0  & \underline{16.8}     &   \textbf{15.5} \\
            bier127      & 15.6    & \underline{10.8}     &  \textbf{6.4}       &pr299           & \textbf{18.2}    & 20.6        &   \underline{18.8} &pr226          & \underline{15.5}  & 18.0     &   \textbf{8.5} \\
            lin318       & \textbf{14.3}    & 16.6        &  \underline{16.3}   &fl417           & 22.7    & \underline{19.2}     &   \textbf{17.3}    &pr439          & 21.4  & \underline{19.3}     &   \textbf{13.7}  \\
           
            \bottomrule
        \end{tabular}
    }
\end{table*}

% \begin{table*}[t!]
%     \caption{The performance of different heuristic methods on the POMO\cite{Yeong2010} algorithm for optimizing the solution to the CVRP. \xl{The results of CAE are the same with those of ReEvo.}}
%     \label{tab:nco_results}
%     \centering
%     \begin{tabular}{lcccc}
%         \toprule
%         \multirow{2}{*}{Method } & \multicolumn{2}{c|}{$n = 500$} & \multicolumn{2}{c}{$n = 1000$} \\
%         \cmidrule(lr){2-3}\cmidrule(lr){4-5}
%         & Obj. &  Gap (\%) & Obj. & Gap (\%)  \\
%         \midrule
%         BASE      & 51.2 & 0.0 & 145.4 & 0.0 \\ 
%         EOH       & 50.7 & 0.9 & 138.0 & 5.0 \\
%         ReEvo     & \textbf{48.2} & \textbf{5.8} & \textbf{127.5} & \textbf{12.3} \\
%         \rowcolor{gray!10}CAE       & \textbf{48.2} & \textbf{5.8} & \textbf{127.5} & \textbf{12.3} \\
%         \bottomrule
%     \end{tabular}
% \end{table*}
%\iclrfinalcopy % Uncomment for camera-ready version, but NOT for submission.

\begin{table*}[t!]
    \centering
    \caption{The performance of different heuristic methods on various algorithms for optimizing the solution to the TSP.}
    \label{tbl:FM}
     \scalebox{0.86}{
        \begin{tabular}{l c ccc c ccc c ccc} 
          \toprule[2pt]
          \multirow{2}{*}{Type}&&\multicolumn{3}{c}{TSP20}&&\multicolumn{3}{c}{TSP50}&&\multicolumn{3}{c}{TSP100}\\
            &&$Obj\downarrow$&$Gap(\%)\uparrow$&$time\downarrow$&&$Obj\downarrow$&$Gap(\%)\uparrow$&$time\downarrow$&&$Obj\downarrow$&$Gap(\%)\uparrow$&$time\downarrow$\\
          \midrule
          GA                    &&6.1&0.0&0.4    &&18.2&0.0&1.3  &&40.8&0.0&2.3\\
          GA+EOH \cite{fei2024eoh}               &&\underline{6.0}&\underline{1.9}&\underline{0.3}    &&\underline{17.8}&\underline{2.3}&\underline{0.8}   &&\underline{40.5}&\underline{0.6}&\underline{2.0}\\
          GA+ReEvo\cite{ye2024reevo}              &&\underline{6.0}&\underline{1.9}&\underline{0.3}    &&17.9&1.3&\underline{0.8}   &&40.6&0.5&2.1\\
          \rowcolor{gray!10} GA+CAE(ours)  
                             &&\textbf{5.7}&\textbf{6.6}&\textbf{0.2}    &&\textbf{16.3}&\textbf{10.3}&\textbf{0.6}  &&\textbf{36.6}&\textbf{10.2}&\textbf{1.3}\\
          \midrule
          ACO                   &&\textbf{3.8}&\underline{0.0}&\textbf{2.1}    &&\underline{5.9}&0.0&\underline{7.6}   &&8.5&0.0&17.9  \\
          ACO+EOH \cite{fei2024eoh}                &&3.9&-0.7&3.5   &&\underline{5.9}&\underline{0.8}&9.1    &&8.5&0.3&17.4\\
          ACO+ReEvo\cite{ye2024reevo}              &&3.9&-0.2&\underline{2.5}   &&\underline{5.9}&0.3&\underline{7.6}    &&\textbf{8.4}&\underline{0.7}&\textbf{12.2}\\
          \rowcolor{gray!10} ACO+CAE(ours)
                                &&\textbf{3.8}&\textbf{0.5}&\underline{2.5}    &&\textbf{5.8}&\textbf{1.8}&\textbf{6.4}    &&\textbf{8.4}&\textbf{1.6}&\underline{13.7}\\
          \midrule
          KGLS\cite{ARNOLD201932}                  &&\underline{4.4}&0.0&\underline{4.1}    &&\underline{6.7}&\underline{0.0}&\underline{10.3}   &&9.3&0.0&\underline{26.8}\\
          KGLS+EOH              &&\underline{4.4}&\underline{0.6}&5.6    &&6.8&-0.2&14.0  &&\underline{9.2}&\underline{0.4}&28.8\\
          KGLS+ReEvo            &&\underline{4.4}&0.2&5.9    &&6.8&-1.0&14.9  &&9.3&-0.3&\textbf{20.9}\\
          \rowcolor{gray!10} KGLS+CAE(ours) 
                            &&\textbf{3.9}&\textbf{11.2}&\textbf{3.5}    &&\textbf{5.9}&\textbf{11.7}&\textbf{9.0}    &&\textbf{8.5}&\textbf{8.2}&28.0\\
          \bottomrule[2pt]
        \end{tabular}
        }
\end{table*}

\subsection{Performance on Benchmark Instances}

Table~\ref{tab:comparisons} highlights the average optimality gap for several benchmark instances, including small-scale and large-scale problems. Across all instances, the CAE framework consistently outperforms the baseline methods, demonstrating its superior optimization capabilities. For example, in the \texttt{ts225} instance, CAE achieves an optimality gap of 4.6\%, significantly lower than GHPP (7.7\%) and ReEvo (6.6\%). Similarly, in the challenging \texttt{rl1889} instance, CAE reduces the gap to 15.8\%, compared to 21.1\% for GHPP and 17.5\% for ReEvo. These results demonstrate the effectiveness of CAE in handling both small and large problem scales.

Moreover, the framework exhibits strong adaptability in solving instances with varying characteristics. In the \texttt{eil51} instance, CAE achieves an impressive gap of 3.5\%, compared to 10.2\% for GHPP and 6.5\% for ReEvo. Even for highly complex instances, such as \texttt{d1655} and \texttt{u1817}, CAE consistently achieves lower optimality gaps, highlighting its robust performance under diverse problem scenarios.

% \subsection{Evaluation on CVRP Instances}

% The CAE framework's performance on the CVRP is presented in Table~\ref{tab:nco_results}. The experiments are conducted for two problem sizes, $n=500$ and $n=1000$, to assess the scalability of the method. For the smaller problem size ($n=500$), CAE achieves a gap of 5.8\%, matching the best-performing baseline method, ReEvo. For the larger problem size ($n=1000$), CAE maintains its superior performance with a gap of 12.3\%, again matching ReEvo while significantly outperforming EOH (5.0\%) and BASE (0.0\%).

% Notably, the results indicate that CAE not only excels in identifying high-quality solutions but also demonstrates scalability and efficiency in solving large-scale optimization problems. The results confirm that the integration of evolutionary optimization and LLM-based code generation within CAE provides a robust mechanism for tackling complex instances of CVRP.

\subsection{Analysis on TSP Instances}

Further evaluation of CAE is conducted on TSP instances of varying sizes (TSP20, TSP50, and TSP100), as summarized in Table~\ref{tbl:FM}. The comparison includes both standard heuristic methods (e.g., GA, ACO, KGLS) and their enhanced variants (e.g., GA+EOH, GA+ReEvo). Across all instances, CAE demonstrates significant improvements in terms of objective value, optimality gap, and computation time.

For small-scale problems such as TSP20, CAE achieves an objective value of 5.7 with a gap of 6.6\%, outperforming all baseline methods, including GA (0.0\%) and GA+ReEvo (1.9\%). Similarly, for medium-scale problems such as TSP50, CAE achieves a gap of 10.3\%, compared to 2.3\% for GA+EOH and 1.3\% for GA+ReEvo. For large-scale problems such as TSP100, CAE achieves a gap of 10.2\%, far exceeding the performance of GA+ReEvo (0.5\%) and GA+EOH (0.6\%).

Additionally, CAE's integration with advanced heuristics such as ACO and KGLS further enhances its effectiveness. For instance, in TSP50, ACO+CAE achieves a gap of 1.8\%, outperforming ACO+ReEvo (0.3\%) and ACO+EOH (0.8\%). Similarly, in TSP100, KGLS+CAE achieves a gap of 8.2\%, significantly improving upon KGLS+ReEvo (-0.3\%) and KGLS+EOH (0.4\%).

\subsection{Insights and Observations}

The experimental results reveal several key insights:
1. Compared to baseline methods, the CAE framework consistently achieves lower optimality gaps, demonstrating its ability to both explore and exploit superior solutions.
2. The two-dimensional optimization approach that integrates structural and functional evolution enhances the framework's ability to discover innovative algorithmic structures.
3. CAE exhibits strong scalability and adaptability, effectively addressing both small-scale and large-scale problems with varying complexity.

These results validate the effectiveness of the proposed framework and underscore its potential to push the frontier in the field of automated algorithm design. The integration of LLMs with evolutionary optimization has given rise to a novel paradigm for solving complex optimization problems, breaking through the limitations of traditional heuristic methods. In addition to these baseline demonstrations, we will continue to provide analyses explaining how our approach expands the scope of using LLMs for algorithm design.
\section{Case Study: Automated Improvement of Large-scale Optimization Methods}\label{sec5}

This section demonstrates how our system autonomously optimizes the solution process for large-scale quadratic optimization problems, with minimal human intervention. We aim to showcase how the proposed framework, through an end-to-end automated pipeline, can completely refine and surpass human-designed algorithms. The case study highlights the ability of the framework to discover non-intuitive solutions and adapt to highly ill-conditioned problems without relying on traditional algorithmic templates or significant human input.

\subsection{LLM-driven Automated Problem Analysis and Solution}

For the quadratic minimization problem:

\begin{equation}
\min_{x \in \mathbb{R}^d} f(x) \coloneqq \frac{1}{n} \sum_{i=1}^n \Big( \frac{1}{2} \langle x, A_i x\rangle + \langle b_i, x\rangle \Big),
\end{equation}

where $A_i \in \mathbb{R}^{d \times d}$ are positive definite matrices and $b_i \in \mathbb{R}^d$ are vectors, the framework begins by analyzing the mathematical structure and properties of the objective function. It then generates problem-specific algorithms tailored to the characteristics of the input data and constraints.

To simulate realistic optimization scenarios, we follow a setup adapted from existing literature. Each $A_i$ is constructed as a diagonal matrix, with elements in the first half sampled uniformly from $[1, 10^{\xi / 2}]$ and the second half from $[10^{-\xi / 2}, 1]$, where $\xi > 0$ controls the condition number of the matrices. The elements of vectors $b_i$ are sampled uniformly from $[0, 10^3]$. This configuration allows us to test the robustness of the generated algorithms under varying levels of ill-conditioning.

The proposed framework employs LLMs to generate optimization algorithm code automatically. The LLM analyzes the problem's mathematical formulation and generates modular, interpretable code that adapts dynamically to the problem's structure. The iterative refinement process integrates feedback from intermediate results to improve algorithmic performance progressively.

For example, a representative template for the LISR-k\cite{liu2024incremental} method is shown below, illustrating the modularity and flexibility of the generated code:

\begin{lstlisting}[language=Python, caption=Core optimization function.]
def search_routine(objective_function: callable, x0: np.ndarray, 
                  A_list: List[np.ndarray], b_list: List[np.ndarray], 
                  max_iter: int = 100, tol: float = 1e-6) -> np.ndarray:
    """
    Core optimization function based on LISR-k method.
    """
    for iter in range(max_iter):
        # Update Hessian approximation
        # Execute optimization iteration
        pass
    return best_solution
\end{lstlisting}

The generated code is modular and extensible, allowing for straightforward incorporation of advanced techniques such as adaptive learning rates, preconditioning, or gradient corrections, depending on the problem's requirements and complexity.

To further improve the initial LLM-generated algorithms, the framework incorporates an evolutionary optimization process. This method systematically explores the algorithm design space to enhance performance in terms of convergence speed, stability, and accuracy. The evolution process is carried out as follows:

\begin{enumerate}
    \item Initialize a population of algorithm variants generated by the LLM.
    \item Evaluate the fitness of each variant based on performance metrics, such as convergence rate, solution quality, and computational efficiency.
    \item Apply mutation (code modifications) and crossover (combination of variants) to generate improved algorithms.
    \item Iterate until convergence criteria are met or performance saturates.
\end{enumerate}

This approach ensures that the framework not only generates initial solutions but also systematically refines them through automated exploration and optimization.

We evaluated the framework's performance on quadratic optimization problems with varying levels of difficulty, characterized by condition numbers controlled by $\xi \in \{12, 16, 20\}$, where higher values of $\xi$ correspond to more ill-conditioned problems. The experiments were conducted with $n = 1000$ and $d = 50$. We compared the convergence rates and optimization accuracy of several methods, including IQN, SLIQN, LISR-1, LISR-k, and LISR-k(CAE).

For $\xi = 12$, IQN and SLIQN exhibited rapid initial convergence but plateaued early, failing to achieve high accuracy. In contrast, LISR-k(CAE) achieved optimal convergence within 50 seconds. As $\xi$ increased to 16 and 20, the performance of IQN and SLIQN deteriorated significantly due to increased sensitivity to ill-conditioning. However, LISR-k and LISR-k(CAE) maintained robust optimization capabilities, demonstrating consistent convergence to high-quality solutions.

Notably, LISR-k(CAE) outperformed all other methods across all scenarios, particularly excelling in high-dimensional and ill-conditioned problems. This highlights the effectiveness of the LLM-based optimization framework, which combines algorithm generation, automated evolution, and adaptive techniques to address challenging optimization tasks.

These experimental results validate the capability of the proposed framework to automatically generate, optimize, and refine algorithms for complex optimization tasks. The LLM-driven approach exhibits exceptional adaptability and robustness, achieving rapid convergence to optimal solutions even under challenging conditions characterized by high condition numbers and high dimensionality. The seamless integration of problem analysis, code generation, and evolutionary refinement underscores the potential of LLMs to revolutionize algorithm design for advanced optimization problems.
\subsection{Model-driven Algorithmic Improvement and Analysis}

\begin{figure*}[ht]
    \centering
    \begin{subfigure}{0.3\textwidth}
        \centering
        \includegraphics[width=\textwidth]{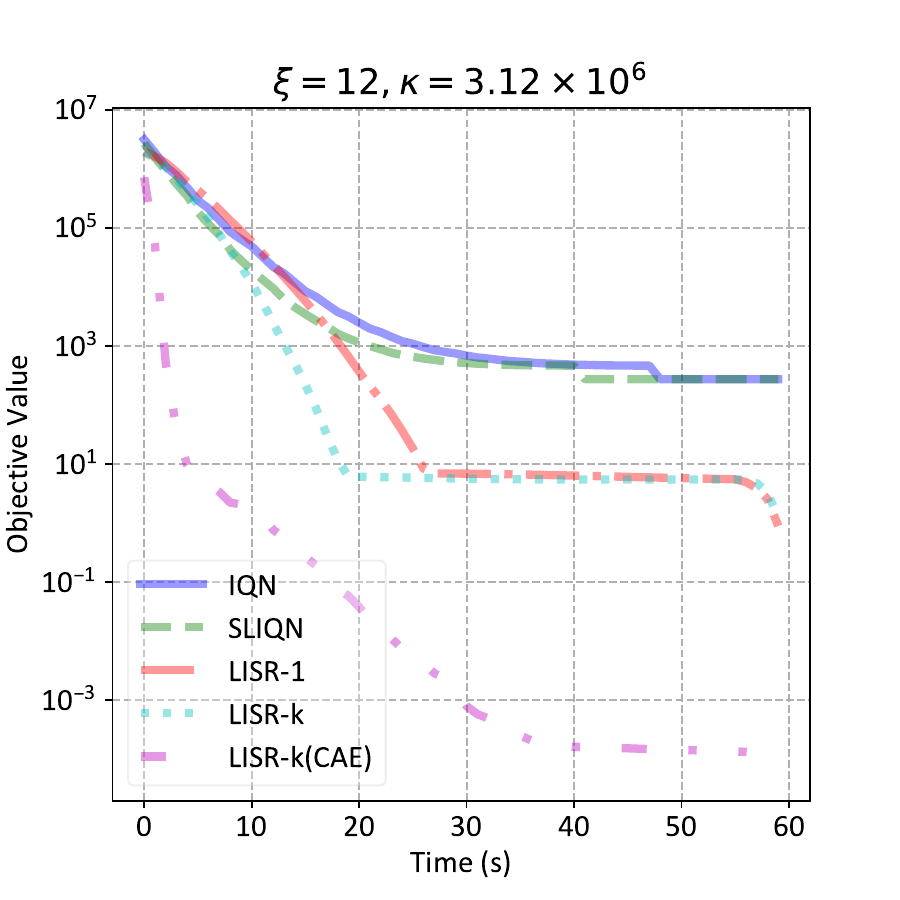}
        \caption{}
        \label{fig1a}
    \end{subfigure}%
    \hfill
    \begin{subfigure}{0.3\textwidth}
        \centering
        \includegraphics[width=\textwidth]{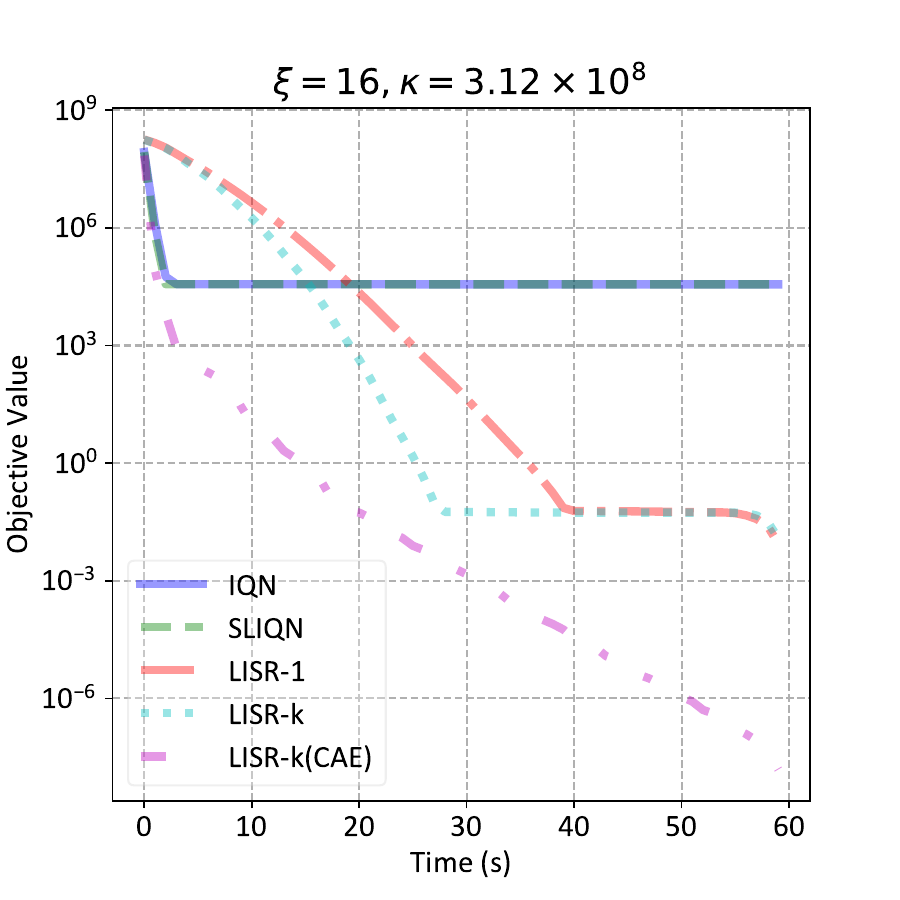}
        \caption{}
        \label{fig1b}
    \end{subfigure}%
    \hfill
    \begin{subfigure}{0.3\textwidth}
        \centering
        \includegraphics[width=\textwidth]{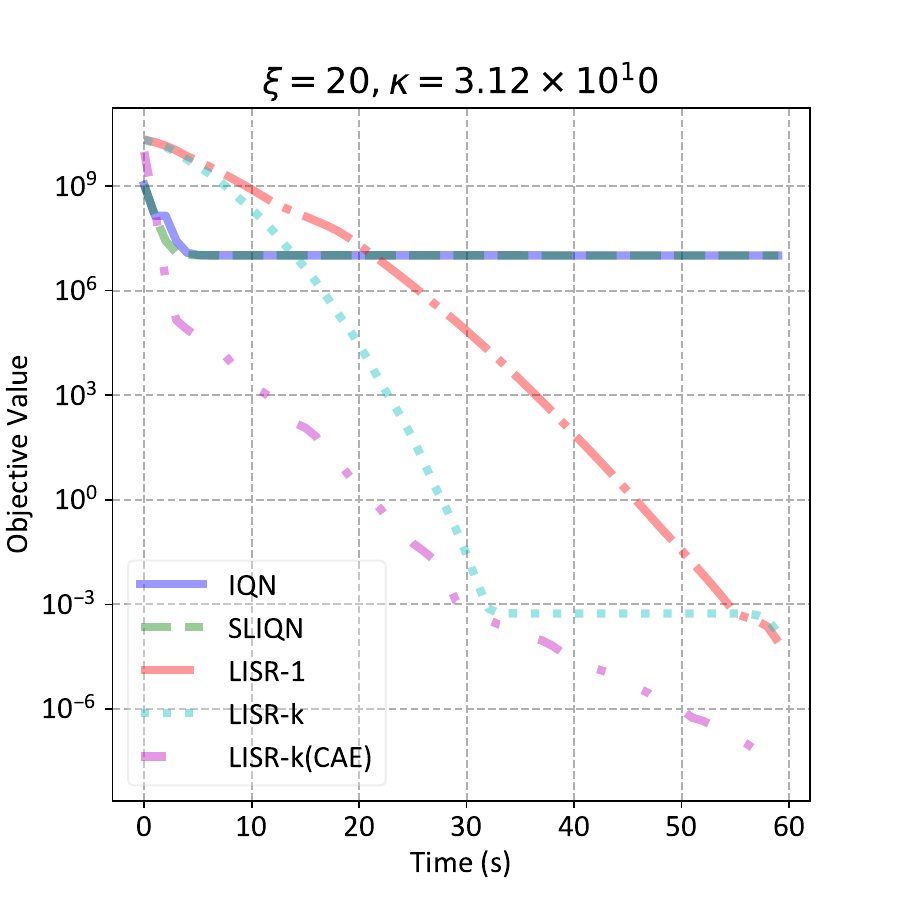}
        \caption{}
        \label{fig1c}
    \end{subfigure}
    \caption{Performance comparison of five optimization algorithms—\textbf{IQN}, \textbf{SLIQN}, \textbf{LISR-1}, \textbf{LISR-k}, and the proposed \textbf{LISR-k (CAE)}—on optimization problems with increasing difficulty (\(\kappa = 3.12 \times 10^6\), \(3.12 \times 10^8\), \(3.12 \times 10^{10}\)). The x-axis represents time (s), and the y-axis shows the objective value. The results highlight that \textbf{LISR-k (CAE)} achieves faster convergence and better performance, particularly in highly challenging scenarios.}
    %\label{fig1}
\end{figure*}
In this section, we investigate the ability of large language models (LLMs) to improve algorithms automatically. We focus on analyzing the optimization process from the baseline algorithm (Algorithm a, i.e., \texttt{iter\_num\_0.py}) to the improved algorithm (Algorithm b, i.e., \texttt{iter\_num\_7.py}) generated by the model. Through function-by-function comparisons and workflow analysis, we demonstrate the contributions of LLMs in code generation, optimization strategy design, and functionality enhancement.

\subsubsection{Function-by-function Comparative Analysis}

To reveal the improvement capabilities of LLMs at the function level, we compare the core implementations of Algorithm a and Algorithm b. The features and shortcomings of Algorithm a are combined with the improvements in Algorithm b to highlight the enhancements and their effects.

\paragraph{Symmetric Rank-k Update (\texttt{srk})}

In Algorithm a, handling singularity issues is straightforward: the rank of \texttt{temp} is checked, and the original matrix \texttt{G} is returned in singular cases. The update depends on direct matrix inversion (\texttt{np.linalg.inv}), which can lead to numerical instability, especially in high-dimensional problems. In contrast, Algorithm b replaces direct inversion with \texttt{np.linalg.solve}, significantly improving numerical stability. It also retains options for pseudoinverse handling in singular cases, offering greater flexibility for further extensions.

\begin{lstlisting}[language=Python, caption=Algorithm a Implementation]
temp = U.T @ (G - A) @ U
if np.linalg.matrix_rank(temp) < U.shape[1]:  # Handle singularity
    return G
return G - (G - A) @ U @ np.linalg.inv(temp) @ U.T @ (G - A)
\end{lstlisting}

\begin{lstlisting}[language=Python, caption=Algorithm b Improvement]
temp = U.T @ (G - A) @ U
if np.linalg.matrix_rank(temp) < U.shape[1]:
    return G
return G - (G - A) @ U @ np.linalg.solve(temp, np.eye(temp.shape[0])) @ U.T @ (G - A)
\end{lstlisting}

\paragraph{Greedy Matrix Selection (\texttt{greedy\_matrix})}

Algorithm a selects rows greedily based only on the diagonal element differences, sorting and taking the top \texttt{k} largest differences. While simple, this approach ignores the overall structure of the matrix, potentially leading to suboptimal choices. Algorithm b improves upon this by using the Frobenius norm to compute row differences, taking into account the global structure of the matrix. This results in more reasonable greedy selection and better global performance for matrix updates.

\begin{lstlisting}[language=Python, caption=Algorithm a Implementation]
diff = np.diag(G - A)
indices = np.argsort(diff)[::-1][:k]
U = np.zeros((G.shape[0], k))
U[indices, np.arange(k)] = 1
return U
\end{lstlisting}

\begin{lstlisting}[language=Python, caption=Algorithm b Improvement]
diff = G - A
row_norms = np.linalg.norm(diff, axis=1)
indices = np.argsort(row_norms)[::-1][:k]
U = np.zeros((G.shape[0], k))
U[indices, np.arange(k)] = 1
return U
\end{lstlisting}

\paragraph{Sherman-Morrison Update (\texttt{sherman\_morrison})}

Algorithm a relies on the classical Sherman-Morrison formula but uses direct matrix inversion (\texttt{np.linalg.inv}), which can be numerically unstable, especially in high-dimensional settings. Singular cases are handled by simply returning the original inverse matrix \texttt{A\_inv}, lacking more sophisticated mechanisms. Algorithm b dynamically chooses between pseudoinverse (\texttt{np.linalg.pinv}) and stable linear equation solving (\texttt{np.linalg.solve}), greatly enhancing numerical stability and adapting better to singularity issues.

\begin{lstlisting}[language=Python, caption=Algorithm a Implementation]
temp = W - U.T @ A_inv @ V
if np.linalg.matrix_rank(temp) < U.shape[1]:  # Handle singularity
    return A_inv
return A_inv + A_inv @ U @ np.linalg.inv(temp) @ V.T @ A_inv
\end{lstlisting}

\begin{lstlisting}[language=Python, caption=Algorithm b Improvement]
temp = W - U.T @ A_inv @ V
if np.linalg.matrix_rank(temp) < U.shape[1]:
    temp_inv = np.linalg.pinv(temp)
else:
    temp_inv = np.linalg.solve(temp, np.eye(temp.shape[0]))
update_term = A_inv @ U @ temp_inv @ V.T @ A_inv
return A_inv + update_term
\end{lstlisting}

\paragraph{Gradient Accumulation and Update}

Algorithm a accumulates gradients and updates the solution \texttt{x} directly but does not adjust the gradient direction, potentially leading to slower convergence. Its update process lacks dynamic adjustment mechanisms, making it more susceptible to noise. Algorithm b introduces a gradient correction term to refine the direction, improving convergence precision. Additionally, a dynamic scaling factor, computed based on the ratio of gradient norms, enhances the algorithm's robustness and adaptability.

\begin{lstlisting}[language=Python, caption=Algorithm a Implementation]
grad_sum = np.sum([np.dot(A_i, z_i) + b_i for A_i, z_i, b_i in zip(A_list, z_list, b_list)], axis=0)
x_new = B_bar_inv @ grad_sum
\end{lstlisting}

\begin{lstlisting}[language=Python, caption=Algorithm b Improvement]
grad_correction = np.sum([np.dot(A_i, x_new) + b_i for A_i, b_i in zip(A_list, b_list)], axis=0)
x_new -= 0.1 * grad_correction / (t + 1)
scaling_factor = np.linalg.norm(grad_correction) / np.linalg.norm(grad_sum)
x_new *= scaling_factor
\end{lstlisting}

Through the above function-by-function comparisons, it is evident that large language models can automatically identify shortcomings in baseline algorithms and propose improvements that enhance numerical stability, structural optimization, and dynamic adjustment strategies. The resulting Algorithm b performs significantly better than Algorithm a, particularly in high-dimensional and complex problems.

\paragraph{\textbf{Workflow Comparison Analysis}}

This section presents a concise comparison of the workflows across three algorithm generations: \texttt{iter\_num\_0.py} (0th generation), \texttt{iter\_num\_3.py} (3rd generation), and \texttt{iter\_num\_7.py} (7th generation). The evolution highlights progressive enhancements in numerical stability, gradient correction, and adaptive scaling, transforming a basic implementation into a robust and efficient optimization framework.
\begin{figure}[h]
    \centering
    \includegraphics[width=0.8\textwidth]{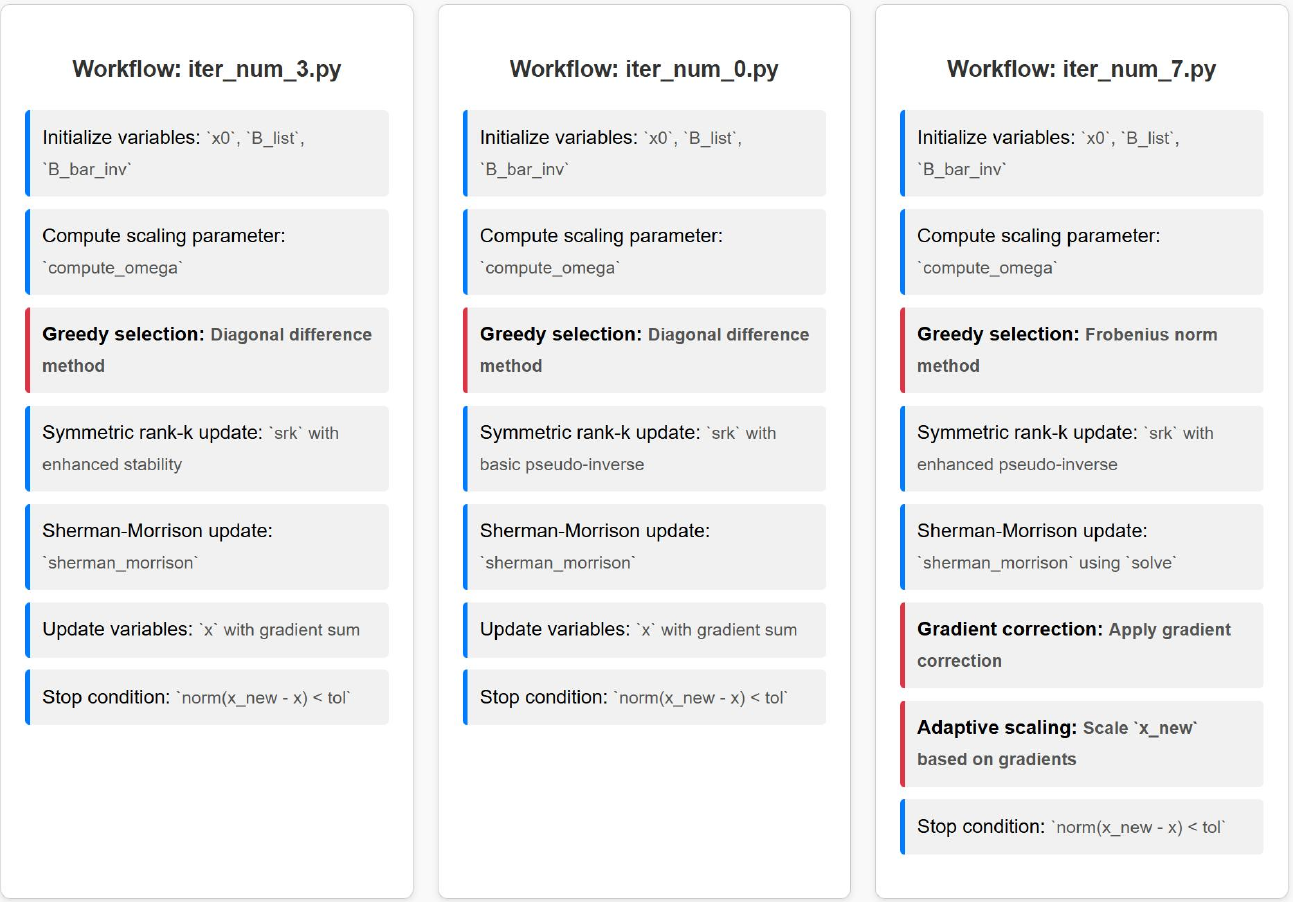} % Replace with your actual file name
    \caption{Workflow comparison across three generations of algorithms: \texttt{iter\_num\_0.py}, \texttt{iter\_num\_3.py}, and \texttt{iter\_num\_7.py}. The \textit{blue regions} represent areas with minor changes, while the \textit{red regions} indicate significant modifications in the workflow. The figure illustrates the evolution from a basic implementation to an enhanced framework, emphasizing improvements in numerical stability, gradient correction, and adaptive scaling strategies.}
    \label{fig:workflow_comparison}
\end{figure}
The 0th generation (\texttt{iter\_num\_0.py}) provides a foundational, straightforward implementation. It initializes the target variable \(x_0\), computes a simple greedy matrix selection based on diagonal differences, performs symmetric rank-\(k\) updates using pseudo-inversion, and updates variables via the Sherman-Morrison formula. While numerically stable for simple problems, its reliance on diagonal sorting and lack of gradient correction limit its performance in handling complex or ill-conditioned problems.

By the 3rd generation (\texttt{iter\_num\_3.py}), significant improvements are introduced. The symmetric rank-\(k\) update now incorporates stability checks (\texttt{np.allclose}) to avoid unnecessary computations, and the Sherman-Morrison update reduces redundant operations, improving both efficiency and numerical robustness. While retaining the basic greedy selection strategy, this version better addresses medium-complexity problems but still lacks advanced selection strategies and gradient correction, which restrict its global optimization capabilities.

The 7th generation (\texttt{iter\_num\_7.py}) achieves comprehensive advancements. It replaces diagonal-based greedy selection with a Frobenius norm-based approach, capturing global matrix characteristics more effectively. Numerical stability is further enhanced by substituting pseudo-inversion with \texttt{np.linalg.solve} in both symmetric rank-\(k\) and Sherman-Morrison updates. Additionally, gradient correction is introduced to refine update directions, while adaptive scaling dynamically adjusts step sizes based on gradient norms, dramatically improving performance in high-dimensional, ill-conditioned problems.

The evolution from \texttt{iter\_num\_0.py} to \texttt{iter\_num\_7.py} demonstrates a clear trajectory of increasing sophistication and applicability. The 0th generation is suitable for simple problems with modest stability requirements. The 3rd generation introduces numerical refinements and is effective for medium-complexity tasks. Finally, the 7th generation integrates advanced selection, correction, and scaling mechanisms, achieving exceptional adaptability and robustness for complex, high-dimensional scenarios.

This progressive innovation in workflow design underscores the strengths of our framework. By combining modular iterative optimization, dynamic greedy selection, and feedback-driven adaptive mechanisms, the framework transcends traditional static workflows. It achieves a versatile, scalable, and highly effective optimization process, capable of addressing a wide range of problem complexities while maintaining structural clarity and functional synergy.

\section{Future Research Directions}\label{sec7}

Despite the significant progress presented in this work, there remain numerous opportunities to further enhance the capabilities of automated algorithm design systems. Below, we outline several key directions for future research:

\begin{enumerate}
    % \item \textbf{Improved Adaptability to Diverse Problem Domains} \\
    % While the proposed framework exhibits strong performance across optimization problems like TSP and CVRP, extending its adaptability to a broader range of domains is a critical next step. For instance, integrating domain-specific knowledge into the LLM-driven process could improve performance in areas such as bioinformatics, robotics, or financial modeling. Additionally, expanding the framework’s ability to handle multi-objective and real-time optimization tasks would further increase its utility across complex, dynamic problem spaces.

    \item \textbf{Enhanced Efficiency and Scalability} \\
    The current framework demonstrates scalability to large-scale problems, but computational efficiency remains a challenge, particularly for problems requiring extensive LLM queries or computationally intensive validation steps. Future work could focus on developing lightweight LLM variants or hybrid methods that combine traditional heuristics with LLMs to reduce computational overhead. Parallel and distributed optimization strategies could also be integrated to accelerate the evolutionary process on large-scale problems.

    \item \textbf{Self-Learning and Continual Improvement} \\
    A promising direction is the development of self-learning systems where the framework continuously updates its knowledge base and optimization strategies based on prior runs. This could involve leveraging reinforcement learning or meta-learning techniques to enable the system to autonomously refine its problem-solving strategies over time, thereby improving its performance on unseen problems.

    \item \textbf{Integration with Scientific Discovery Pipelines} \\
    Beyond algorithm design, this framework could be extended to automate broader scientific discovery pipelines. For example, integrating it with experimental design, data analysis, and hypothesis generation workflows could significantly accelerate the pace of scientific research. Such systems could serve as collaborative tools for researchers, autonomously generating insights and solutions that augment human creativity and expertise.

    \item \textbf{Exploration of Explainable AI in Automation} \\
    As automated systems increasingly influence algorithm design and scientific exploration, ensuring transparency and interpretability becomes essential. Future work could focus on developing explainable frameworks that not only generate optimal solutions but also provide clear, interpretable insights into the reasoning behind those solutions. This would foster greater trust and usability in scientific and industrial applications.

    \item \textbf{Ethical Considerations and Safe Automation} \\
    As the automation of algorithm design advances, ethical considerations must also be addressed. These include ensuring that the generated algorithms are unbiased, fair, and safe for deployment across various applications. Future research could explore mechanisms for embedding ethical principles directly into the optimization process to ensure responsible and sustainable innovation.
\end{enumerate}
\section{Conclusion}\label{sec8}

This paper presents a novel framework for algorithm design that achieves fully automated and innovative algorithm generation through global design exploration, holistic optimization, and autonomous search space expansion. Extensive experiments and cross-domain case studies have thoroughly validated the framework’s outstanding adaptability, scalability, and superior performance in addressing complex computational problems, further demonstrating its immense potential to disrupt traditional algorithm design paradigms and reshape computing system architectures.

Our work represents a decisive stride toward fully automated algorithm design, significantly lowering the barrier for developing high-performance algorithms while offering fresh solutions to the bottlenecks caused by manual design and local optimization in the past. By confronting the multifaceted challenges of scalability, adaptability, and structural innovation, this paper proves that automated algorithm design can not only replicate the expertise of human specialists but also transcend existing knowledge boundaries to discover unprecedented innovations.

These advancements inject new momentum into accelerating computational research and scientific exploration, indicating that future automated systems will play an increasingly crucial role in driving technological breakthroughs across interdisciplinary fields, optimizing complex engineering systems, and exploring uncharted areas of computation. With this breakthrough, our framework lays a solid foundation for the next generation of intelligent systems and heralds the dawn of a new era in autonomous algorithm discovery.
\bibliography{sample}
\bibliographystyle{IEEEtran}
%\printbibliography

\end{document}